\documentclass[a4paper,11pt]{article}
\pdfoutput=1 

\usepackage{jinstpub} 

\usepackage{lineno}

\title{\boldmath Optimisation of Operating High Voltage of the large area Resistive Plate Chamber for ICAL experiment}

\author[a,1]{S. Pethuraj,\note{Corresponding author.}}
\author[a]{G. Majumder,}
\author[a]{K.C. Ravindran,}
\author[a]{B. Satyanarayana,}
\author[a,b]{Umesh L}

\affiliation[a]{ Tata Institute of Fundamental Research,\\Mumbai-400005, India}
\affiliation[b]{The American College,\\ Madurai-625002, India}

\emailAdd{s.pethuraj@tifr.res.in}

\abstract{The Resistive Plate Chamber is a widely used detector in high energy physics.
The operating potential of this chamber is determined by the optimisation of the efficiency
and noise rate of the device. This optimisation is based on the assumption that the
performance of the device over the whole surface area is uniform. The INO-ICAL experiment
is going to use $\sim$ 30000  RPC of size $\sim$2\,m$\times$2\,m. All the RPC will have to
pass a minimum quality assurance criteria, but may not be able to maintain a good
uniformity over the whole surface area, particularly for the whole running period
of about twenty years. This paper describes the choice of the optimum
operating HV for an RPC of non-uniform response.
}


\begin{document}
\maketitle
\flushbottom
\section{Introduction}
\label{chap:introduction}  
The proposed India-based Neutrino Observatory (INO) at Bodi hill
($9^\circ 58'N, 77^\circ 16' E$) in the Southern part of India, with
at least 1\,km rock overburden in all directions will be
the largest experimental facility of basic science
in India to carry out front-ranking experiments in the field of
particle and astroparticle physics.
The Iron Calorimeter (ICAL) detector is the proposed underground neutrino
physics experiment
in the INO cavern, which aims to observe the neutrino oscillation
pattern at least over one full period.
Main goals of this experiment
are the precise measurements of neutrino oscillation parameters
including the sign of the 2-3 mass-squared difference,
$\Delta m_{32}^2 (= m^2_3 - m^2_2)$ through matter effects,
and, last but not the least, the search for any
non-standard effect beyond neutrino oscillations \cite{ino_pramana}.
The detector size is
48\,m\,$\times$\,16\,m\,$\times$\,14.5\,m made of 151 layers of
5.6\,cm iron plates interleaved with the Resistive Plate Chamber (RPC)
for the detection of the muon trajectory. There will be $\sim$ 30000 RPCs
of size $\sim$2\,m\,$\times$\,2\,m, made of glass gap with pickup
panels and corresponding electronics to collect and store the induced signals which are produced due to the avalanche of electrons along
the trajectory of the charged particles.
Initially, a small number of this large RPC \cite{santanico} were built in the laboratory in
collaboration with the industry. After establishing the whole procedure, jobs are
given to different industries, e.g., production of the gas gap, pickup panel, electronics,
power supply etc., are given in different companies. After assuring the quality control (QC)
of each component separately, all are brought back to the laboratory for final assembly and
test the performance of the RPC. Though there is a stringent condition on the QC at
the vendor's place, the performance of the finally produced RPC is slightly worse than the earlier
produced RPC in the laboratory. One example is the uniformity of the gas gap and consequently
the variation of gain/noise rate of RPC as a function of the position in the RPC. To have
a higher gain of the RPC in the whole surface area, the noise rate, as well as the probability of
streamer formation, is also increased in the high gain area of the detector, which eventually deteriorates the
position and timing performance of the RPC detector. The optimal choice of the applied HV is
obtained by considering the efficiency and resolutions. The outline of the paper is the
following. Section \ref{chap_exptsetup}
describes the experimental setup. Data analysis techniques are described in
Section \ref{chap_analysis}. Section \ref{chap_conventional} shows the overall efficiency of a poor
quality RPC as a function of HV and the V-I curve of the
detector. Section~\ref{chap_multiplicity} discusses the response of
the RPC in different regions of the detector area. The overall
efficiency is calculated based on the optimum selection criteria explained
in Section~\ref{chap_efficiency}. The section
\ref{chap_resolution} shows the position and time resolution as a function of HV and finally
we conclude our findings in section \ref{chap_conclusion}.

\section{Experimental Setup}
\label{chap_exptsetup}
RPCs were introduced in the 1980s in the particle physics experiments as
a replacement for spark counters. Due to its rugged structure and
its large area coverage with uncompromised efficiency and time
resolution made it a significant element in the High Energy
Physics as Trigger detector and Time-of-Flight detector in various experiments. 
The operation of the RPCs are based on the detection of the gas
ionization produced by the charged particle when it passes through the active area of the
gaseous chamber and subsequently the avalanche due to the strong electric field
in that region. The RPCs used as trigger detectors have a
gas gap in the order of a few mm. If the RPCs are used in the Time of Flight
(ToF), the gap in the order of a few 100\,$\mu$m used to get the better
time resolution (few 10 of ps). The gas mixture along with an operating
voltage decides the working mode of the detector whether it is
an avalanche or streamer. In avalanche mode, the electron
multiplication occurs due to the drifting and collision of the
electrons with gas molecules. The avalanche became the precursor of
the new process called streamer if the electron multiplication
reaches the extreme value \cite{raether,meek}.
In the Streamer phase, the kinetic energy of the electron is low and
the recombination of the electron-ion results in the photons and this
create delayed secondary avalanches with respect to the first
one. If the number of photons are large enough and the applied
electric potential is strong enough will produce a large number of
the secondary avalanches until the local density and electron-ion
distributions are such to create a plasma between two electrodes (electron and ion
distribution connects the two electrodes). In this process, the
extremely high current such as 100 time larger than the avalanche is
flown through the gas gap, until the electrons and ions are collected
by the electrode.

The RPCs used in this study are made by placing two thin glass plates of 3\,mm thickness, 2\,mm apart from
each other. The gap between the two glass plates is maintained to be 2\,mm using
poly-carbonate "button" spacers. The sides are sealed with side spacers to form a closed chamber.
There are gas inlet and outlet nozzles at the corners of the chamber through which the mixed
gas is flown. The outer surfaces of the glass chamber are coated with a thin film of graphite
paint for the application of external high voltage. The resistivity of the coat is
chosen by optimising the application of uniform
electric field in the gap as well as localising the induced signal in a small area.
A thin layer (50\,$\mu$m) of insulated mylar sheet is kept in between the chamber and
pickup panels on both sides. The pickup panels are made of the copper strips with a
width of 2.8\,cm and an inter-strip gap of 0.2\,cm pasted on the plastic honeycomb
material, where the other side is pasted with a thin aluminium layer for grounding.
The RPCs are operational in avalanche mode with a non-flammable gas mixture of
$\mathrm{C_2 H_2 F_4}$ (95.2\%), iso-$\mathrm{C_4H_{10}}$ (4.5\%) and $\mathrm{SF_6}$ (0.3\%), which are
continuously flown through the gas gap with a rate of 6SCCM per RPC.

To work on the RPCs in avalanche mode, the major component of
the gas mixture is the electronegative freon gas ($\mathrm{C_2 H_2 F_4}$) with enough primary
ionization production along with that low pathlength for the
electron capture. Due to the electronegative property of the gas, it
controls the electron multiplication within the Meek limit \cite{raether,meek}. The
iso-$\mathrm{C_4 H_{10}}$ is the polyatomic gas, which has high absorption
probability for the UV photons produced in the electron-ion
recombination. The absorbed energy is dissipated as the
rotational-vibrational energy levels. The third gas $\mathrm{SF_6}$ is used
with minimal volume fraction but it plays a dominant role in operating the
RPCs in the avalanche mode by quenching the electrons.

Few RPCs with a large non-uniformity in terms of efficiency and noise are studied to find the
optimum operating HV. There are many sources of the non-uniformity, e.g., uniformity
of the thickness of the glasses \& supporting spacers to maintain the gas gap, resistivity
in the graphite coating, flatness of the pickup panels, difference in the gain of
amplifiers as well as the threshold of the discriminators, the characteristic impedance of the
pickup strips due to variation of width as well as depth of the supporting material etc.

\begin{figure}
  \center
  \includegraphics[width=0.75\textwidth]{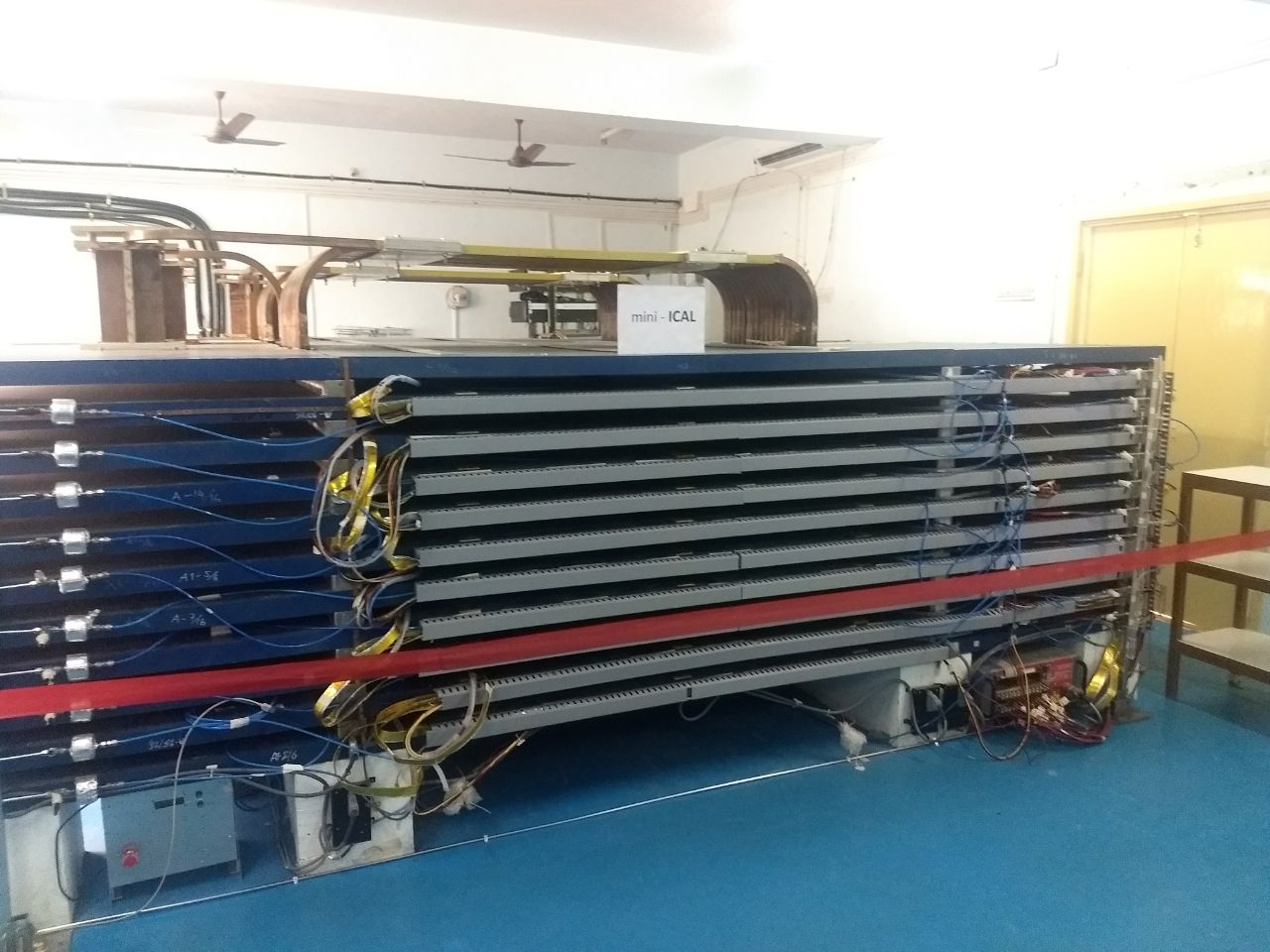}
  \caption{A view of 85\,ton miniICAL detector with 10 layers of RPC detector.}
  \label{fig:miniical_view}
\end{figure}

RPCs are tested using the existing 85\,ton miniICAL experimental
setup \cite{miniical_iichep}, which is a miniature version of ICAL and a schematic view
is shown in figure \ref{fig:miniical_view}. In absence of easily available test beam
facilities with muon beams, the miniICAL is used to trigger cosmic muon trajectory. Though
the miniICAL performance is being tested with the 1.4\,T magnetic field, this study used the
data collected without any magnetic field. Also, there are 20 RPC (two RPC in a gap
with 10 RPC in a vertical column) in the system, but this study is based on 10 RPC of
a column \footnote{The numbering of RPCs starts from the bottom and the top layer is numbered as 9.}.

The characteristic impedance of the strip is around $\sim$50\,$\Omega$. The
readout strips are placed orthogonally on either side of the RPC to measure the position
of the traversed charged particle (X-coordinate from the bottom strip and Y-coordinate from
the top strip plane). The induced signals due to the avalanche multiplication
inside the RPC are amplified and discriminated by an 8-channel NINO ASIC chip \cite{nino}.
The LVDS output of NINO is fed to the FPGA based RPC DAQ system. Finally the electronic
outputs
of an RPC are 128 logic units, to identify the induced signal in the strips and sixteen time
measurements. The 128 logic unit is defined as hit\footnote{Induced signal in a strip larger
 than 100\,fC.} and used to localise the position of the traversing charged particle.
For the time measurement, every 8$^{th}$ strip are connected together to have a single output
of 8 strips. 

This study is performed using cosmic muon events, which are triggered by the coincidence of signals in four fixed RPCs in
that column with the time window of 100\,ns. The logical ``AND'' is done for X- and Y-plane independently to form the trigger signals from X- and Y-plane. The final trigger is created by logical ``OR'' of the trigger signals from X- and Y-plane.  

\section{Data Analysis}
\label{chap_analysis}

One of the main disadvantages of this setup
is the uncertainty of the position and time measurement of muon trajectories in the test RPC.
In test beam setup, muon position is measured with the tracking device of position resolution
about $\sim$mm or less and time is measured very precisely from the beam delivery
system \cite{cmsrpc}. In this setup, those are measured in a crude way.

The typical noise rate of a strip is $\sim$100\,Hz at 10\,kV. The four-fold coincidence logic reduces the
events triggered by the random coincidence of noise. But, muon is contaminated by the
correlated electronic noise as well as cosmic showers mainly due to interactions of remaining
hadrons of cosmic ray particles on the roof of the building and the iron in miniICAL.

Most of the analysis framework is done independently using the information in X-strips and Y-strips
of different layers independently and then combined together to localise the position of the muon
in two dimensional space in an RPC.

The muon position in the RPC is
simply the mean of the strip positions in the RPC. For the timing measurement, strips with
an earlier timestamp amongst different strips are used as the measured time for that layer.
To remove the background events, each layer is considered only when there are at most three
hits in a plane and all three hits should be in consecutive strips. This criterion is used also
to reduce the small fraction of streamer pulse. Position resolution of an RPC is drastically
deteriorated with the multiplicity 4 or more \cite{pethu1} and will be discussed more in
section \ref{chap_multiplicity}.

The position alignment of an RPC with respect to the remaining system is done through an
iterative procedure, where the muon trajectory is fitted with
a simple straight-line equation using the information of all other detectors.
The test layer is aligned by comparing the measured position in that layer with respect to
the extrapolated position from the fit using other layers. This entire process is repeated iteratively where residual distributions of
positions are corrected and updated for all layers.
The details of the alignment technique are given in \cite{sumanta1} and the same iterative
 procedure is also used for the correction of time offset, but for the timing in a layer
there are sixty-four offset corrections (for each and every individual strip) per plane.

After all these criteria, straight lines for the muon are fitted in both X-Z plane and Y-Z
plane independently. Each fit must contain at least six layers and $\chi^2/ndf$ is less than
two are considered for further study. Applying these conditions simultaneously on both X-Z
and Y-Z plan select almost 100\% pure muon signal in the event.
Using the fit parameters, muon trajectories are
extrapolated to the test RPC. The position resolution
of the RPC's are $\sim$8\,mm \cite{pethu1}. Thus the error on the extrapolation on top of the
miniICAL is also about $\sim$6-7\,mm, but this is minimum in the middle layers, which is
about $\sim$3\,mm. Similarly, the uncertainty of the time measurement in each layer is
$\sim$1\,ns \cite{apoorva} and error on the extrapolated time at the top of miniICAL is
0.8-0.9\,ns, whereas at the centre of the stack it is $\sim$0.4\,ns. These extrapolated uncertainties
on position and timing are larger than the typical values achieved in test beam
facilities \cite{hcaltestbeam}. Measured position and time resolutions are convoluted with the
error due to extrapolation, but for the comparison of the performance as a function of applied HV,
the effect will be the same for all HV. The choice of optimum HV will not affect much with
these extrapolated errors.
   
\section {Efficiency Plateau of the RPC detectors}
\label{chap_conventional}  
\begin{figure}
  \center
  \includegraphics[width=0.99\linewidth]{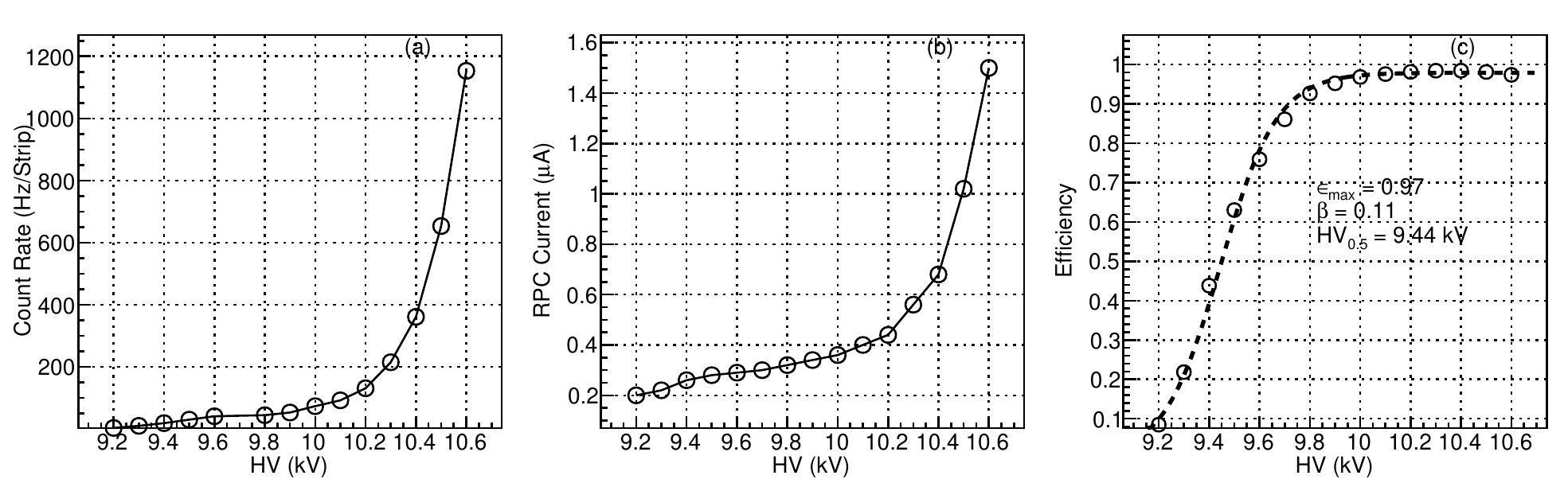}
  \caption{(a), (b) and (c) are average count rate per strip, the V-I curve and the measured efficiency vs high voltage for the RPC under observation.}
  \label{fig:effvshv}
\end{figure}

During the preliminary testing of the RPC under the study, the first
parameter is to measure the Voltage-Current characteristics of the RPC.
The measured RPC count rate per strip and current at different HV is
shown in Figs.~\ref{fig:effvshv}(a) and (b) respectively. 
In general, the choice of the operating high voltage (HV) of the RPC detector is determined
by the overall detection efficiency of the RPCs for the charged particle 
as a function of applied HV. The operating point (workingpoint) of the RPCs are decided by the ``knee'' voltage; the voltage at
which the overall efficiency of the RPCs reached 95$\%$ of the
maximum efficiency\cite{cmsrpc1,argoybj1}. 
The efficiency of the RPC discussed in this section is estimated for the
pixel\footnote{Area of 3\,cm$\times$3\,cm in the RPC detector to match the strip pitch.} when
muon passes through the pixel. The size of the pixel is 3\,cm, whereas the extrapolation
error $\sim$8\,mm, to incorporate the uncertainty of extrapolation, any hit in the RPC strips
with the 3\,cm from the extrapolated position is used for the efficiency measurement.
Even with the disadvantage due to error in track extrapolation, there is also an advantage over
 test beam or conventional muon telescope.
In general, the detection efficiency of the selected strips is estimated using the
cosmic muon telescope made of scintillators as trigger detectors and RPCs as 
detectors under test. In that method, the performance of only a selected
region of the RPC is studied. The same is true for test beam activities also. But to study
the detailed (pixel-wise) performance of the RPC detectors over the whole area, the RPC
detectors kept in miniICAL or any other RPC stack is certainly a better option. It was
mentioned earlier that the uncertainty of the extrapolation is minimum in the middle of
the stack, but due to other considerations, the HV scan measurement is performed for the
RPC, which is kept in the 8th layer of the miniICAL. Muons events are collected with the
in-situ trigger from the time coincidence of layers 2, 3, 4 and 6.
In the different runs, the HV and all other conditions of all RPCs are kept the same except
the HV of layer 8. The muon trajectory is fitted with the hit points of all layers
 excluding layer-8 and interpolated to layer-8 and matched with the measured hit position
 in that layer.
The overall efficiency as a function of High Voltage for the RPC under observation is given
Fig.~\ref{fig:effvshv}(c). From the overall
 efficiency measurement it is observed that the
the efficiency of the RPC is saturated at around 10.2\,kV of the applied
high voltage. The efficiency as a function of the HV is fitted using
the sigmoidal function is given in Eqn~\ref{eqn:sigmoid}\footnote{The observed behaviour of the X- and Y-plane is similar. Thus the results from one of the planes is described in further sections.},
\begin{equation}
 \centering
 \epsilon = \frac{\epsilon_{max}}{1+exp[-(HV-HV_{0.5})/\beta]}
\label{eqn:sigmoid}
\end{equation}
where $\epsilon_{max}$ is the maximum efficiency as HV is
$\infty$. $HV_{0.5}$ is the $HV$ value for which the efficiency is
half of the maximum efficiency ($\epsilon_{max}$). 

\section{Pixel-wise efficiency and Strip Multiplicity for different HV}
\label{chap_multiplicity}
\begin{figure}
  \center
  \includegraphics[width=0.99\linewidth]{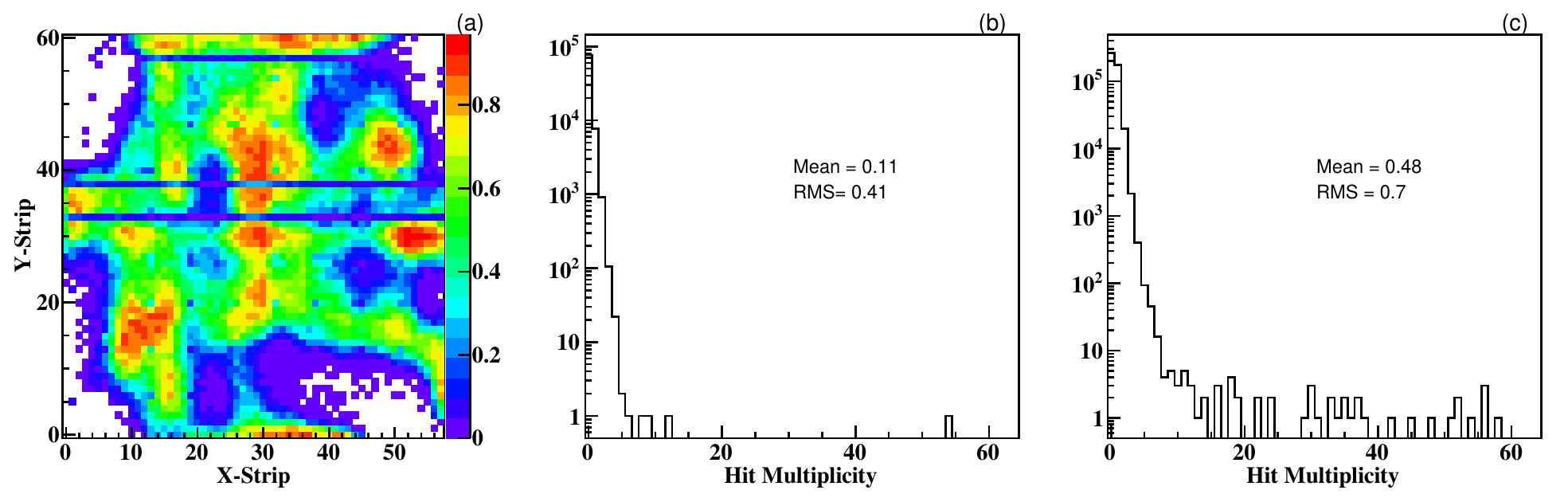}
  \caption{(a), (b) and (c) are the pixel-wise efficiencies and hit multiplicity distribution for efficiency zone 20-30\,$\%$ and 50-60\,$\%$ of the RPC at applied voltage of 9.4\,kV.}
  \label{fig:effhitmult9.4kv}
\end{figure}

Comparing the knee of the efficiency plots with other results, e.g., Ref.~\cite{cmsrpc}, the large
difference is observed between the applied HV for the saturation in the efficiency and the 50\% efficiency
of the RPC. This behaviour might be due to the variation of gain in different regions of the RPC,
where certain regions have larger gain and reach the maximum efficiency earlier than the regions with low gas gain.
The figure \ref{fig:effhitmult9.4kv}(a) shows
the pixel-wise combined efficiency of X\&Y-side for the whole RPC at 9.4\,kV. The three
bands parallel to the Y-axis are due to the malfunction of electronics in the three strips. There
are regions where efficiency is almost zero and some regions are more than 90\% efficient.
The efficiency will increase with applied HV, but to have saturated efficiency of these
low gain regions, one needs a substantial increase of applied HV.

Depending on the efficiency in the RPC at 9.4\,KV, the whole surface area of the RPC
is divided in 10 zones,
with efficiency, 2-10\%, 10-20\% and so on.

Along with the overall efficiency of the RPCs, the other parameter which tells about the
behaviour of the RPC detectors is the cluster size or hit multiplicity distribution.
The hit multiplicity distribution for the regions with efficiency of 20-30\,$\%$ and 50-60\,$\%$ at the applied voltage of 9.4\,kV are
shown in Figs.\ref{fig:effhitmult9.4kv}(b) and (c). In general for a stable operational
RPC, the strip
multiplicity produced by avalanche due to the passage of the muon through RPC detector
is upto three strips. The hit multiplicity beyond three and until $\sim$20 is mostly due
to the hadronic shower and steamer development inside the gas gap. The events with strip
multiplicity of more than $\sim$20 is due to electronic noise that occurred due to EMI from various
sources around the experiment. But, there is no sharp boundary between the streamer and
electronic noise. From the operational point of view, hit multiplicity increases with the applied HV.

\begin{figure}
  \center
  \includegraphics[width=0.45\linewidth]{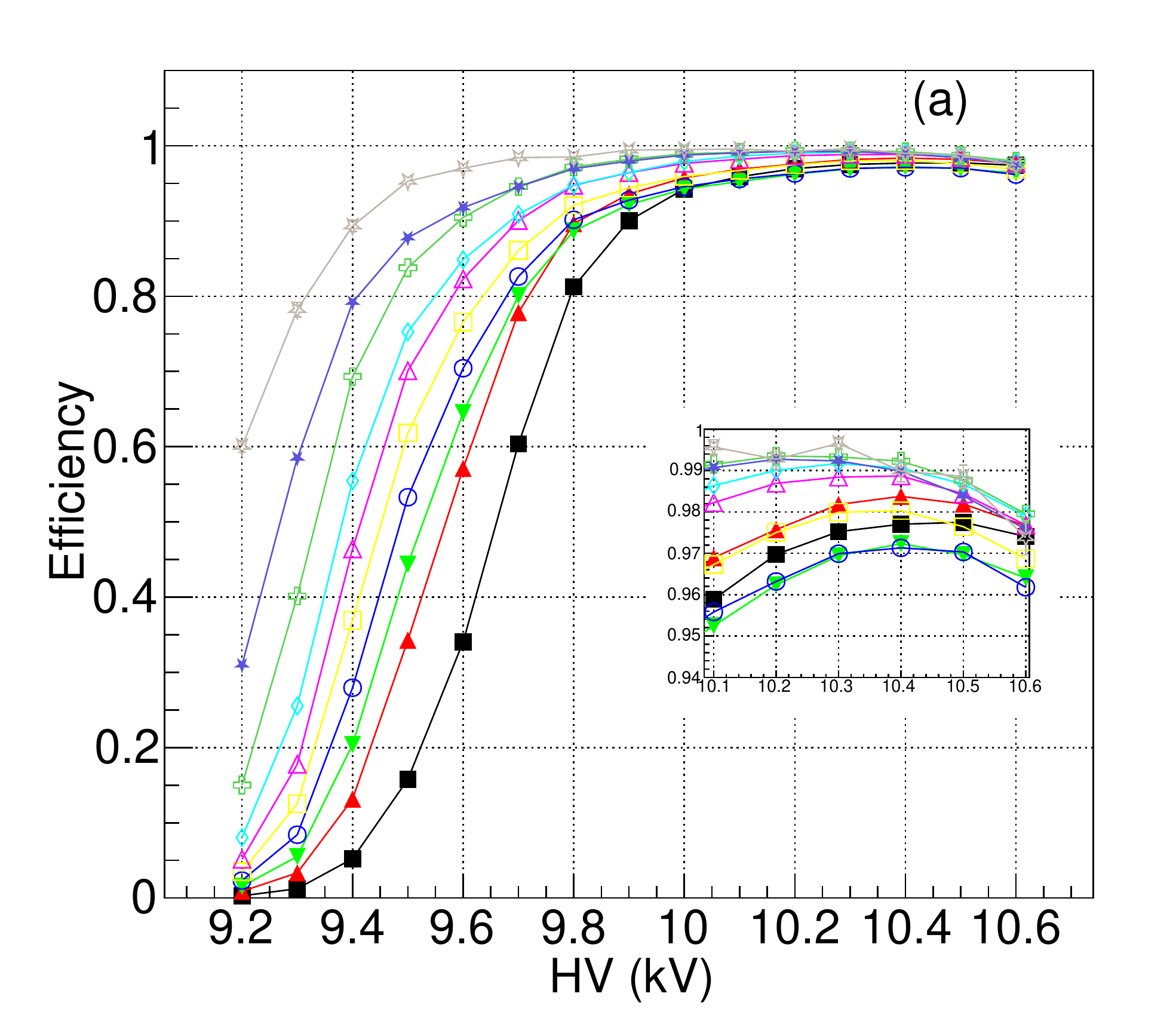}
  \includegraphics[width=0.45\linewidth]{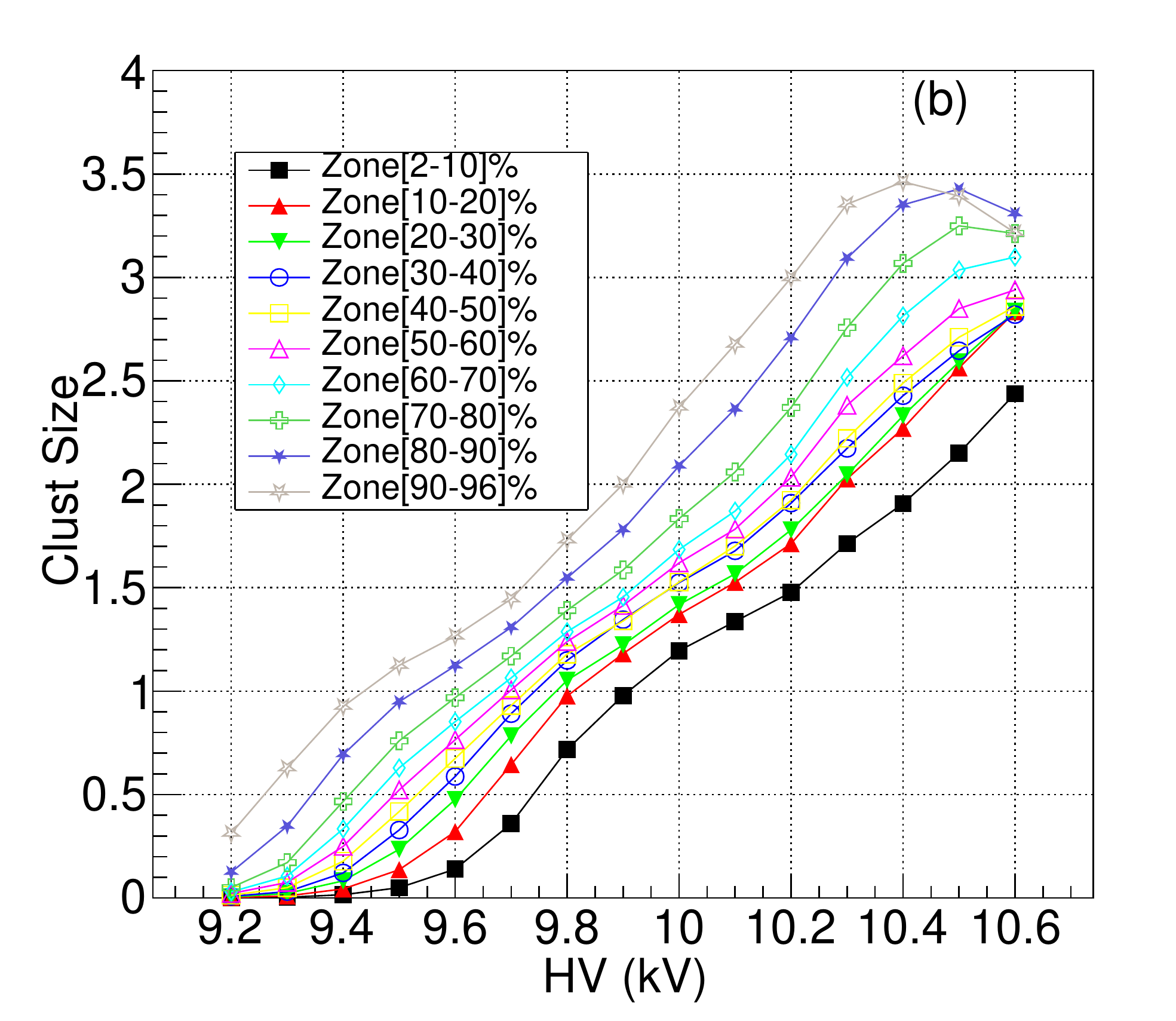}
  \caption{(a) and (b) are the measured efficiency and average cluster size vs high voltage for different zones of the test RPC.}
  \label{fig:effvshvrange}
\end{figure}
\begin{figure}
  \center
  \includegraphics[width=0.99\linewidth]{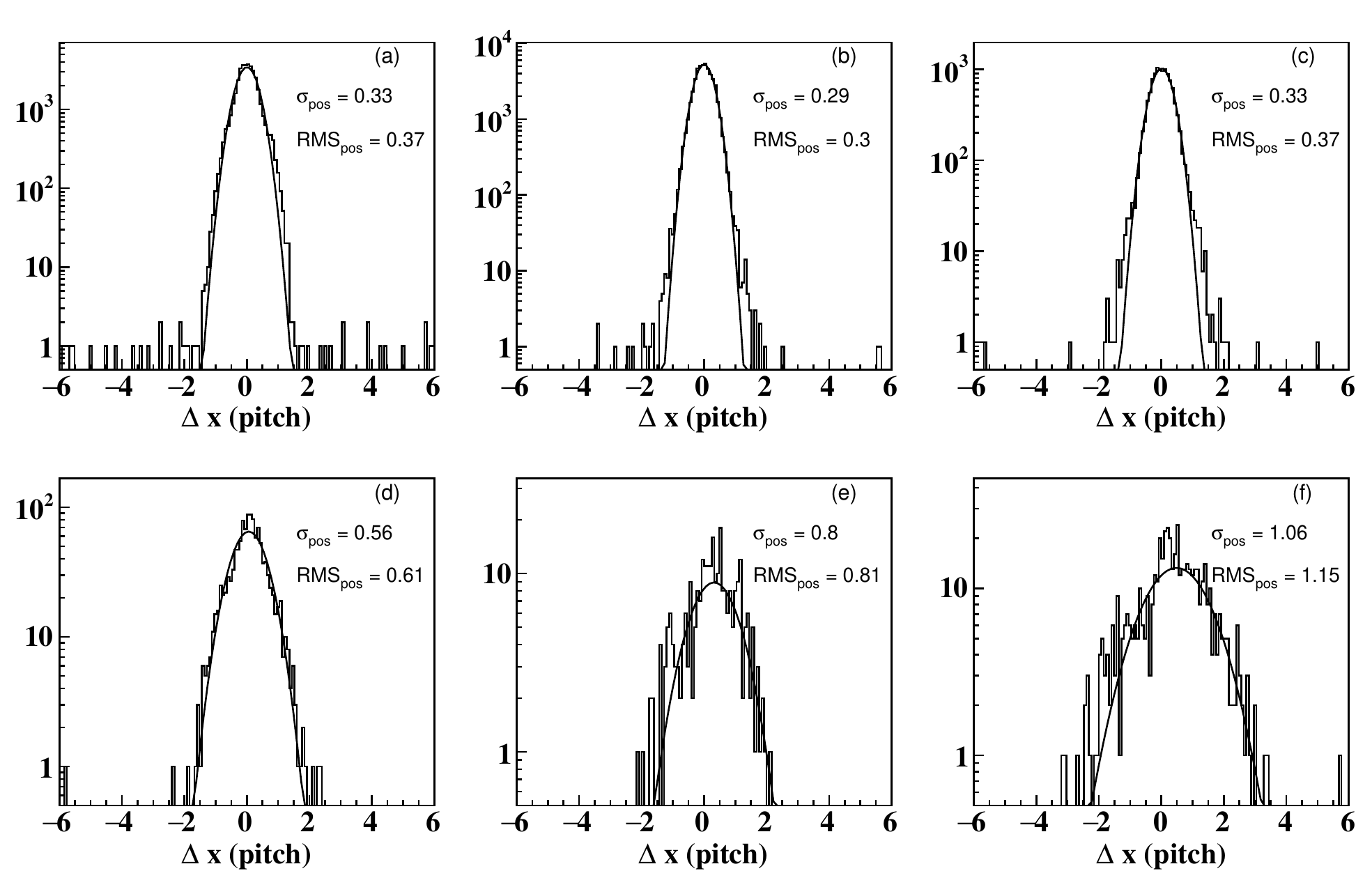}
  \caption{The measured position resolution of the observation layer for different multiplicity from (a) one to (e) five and (f) greater than four at the 60-70\,$\%$ efficiency region.}
  \label{fig:pos_reso_mult12345}
\end{figure}
\begin{figure}
  \center
  \includegraphics[width=0.99\linewidth]{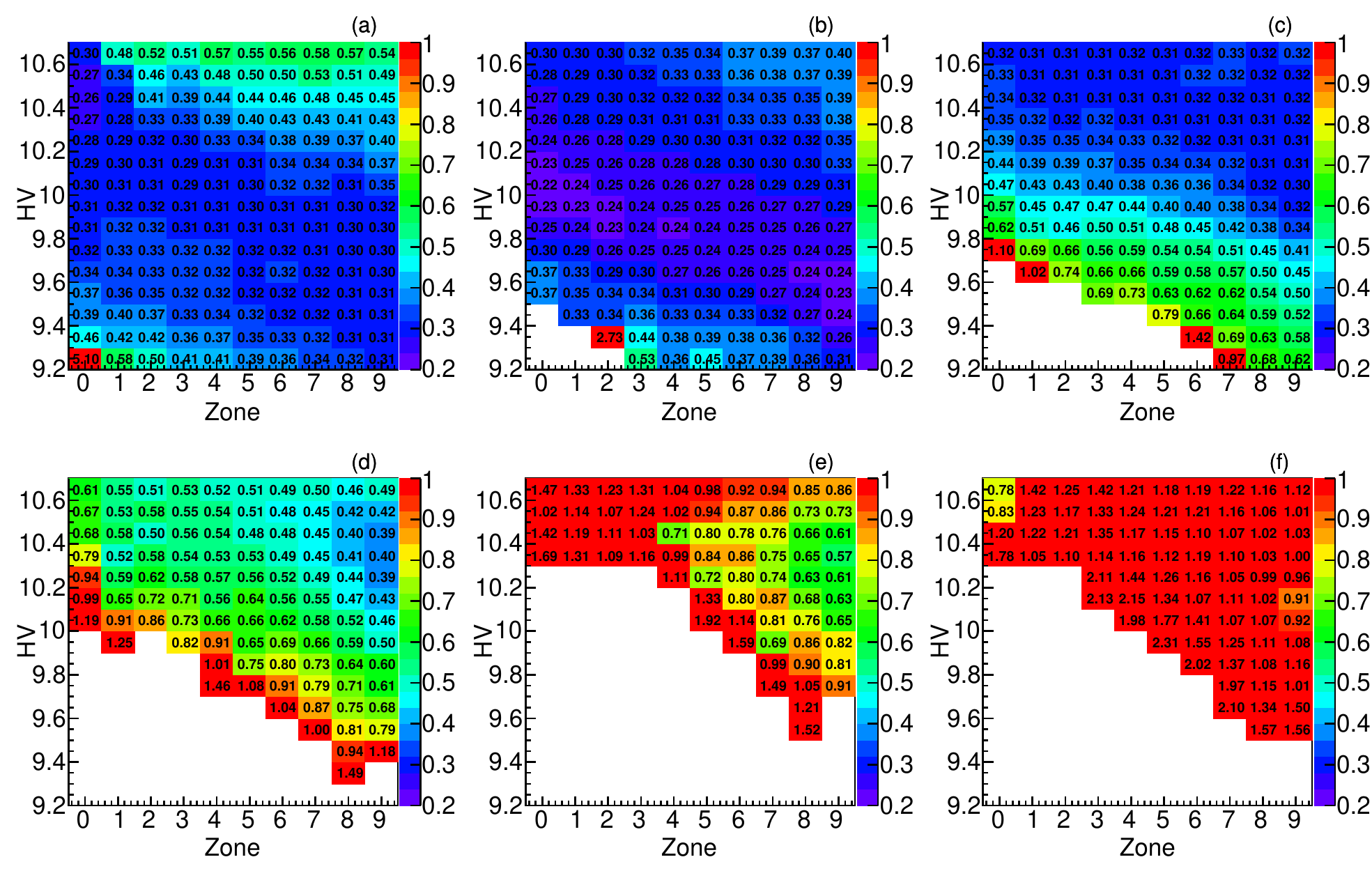}
  \caption{The measured position resolution of the observation layer for HV vs efficiency region for different multiplicity from (a) one to (e) five and (f) greater than four.}
  \label{fig:pos_reso_mult12345_2d}
\end{figure}

The average efficiencies
in one of the RPC planes at different HV
for different zones are shown in the Fig.~\ref{fig:effvshvrange}(a). It is clearer from the
figure that the different zone reaches the saturated efficiency at different HV. The inset plot
of Fig.~\ref{fig:effvshvrange}(a) shows the decrease of efficiency at very high voltage, which is
due to a visible dead time of the detector caused by the large noise rate.
The average hit multiplicities (cluster size) for
different HV at various efficiency zones are shown in Fig.~\ref{fig:effvshvrange}(b). Similar to the efficiencies, the multiplicities are also a
combined function of the zone and applied HV. It was mentioned earlier that one of the sources
of high multiplicity is the streamer formation and it causes the deterioration of position resolution.
Earlier studies \cite{pethu1} showed that the position resolution drastically deteriorates
with the multiplicity of four or more. The operation of this RPC at 10.1\,kV will have a
large region with an average multiplicity of four or more and can not be used for any physics study,
e.g., identification of trajectories and the accurate measurement of momentum
due to poor position resolution. The position resolution for different hit multiplicities from one to five and multiplicities greater than four hits for the observation RPC at 10.1\,kV for efficiency region of 60-70\,$\%$ are given in Figs.~\ref{fig:pos_reso_mult12345} (a) to (f). The observed position resolution for applied HV vs efficiency zones for different multiplicities (from one to five and greater than four) are given in Figs.~\ref{fig:pos_reso_mult12345_2d} (a) to (f). Some regions are empty due to the omission of resolutions with a very low number of events.

At the low HV, events with larger multiplicities, e.g., three or more are dominated by noise and similarly at larger HV, events with
low multiplicity are also dominated by noise associated with inefficiency of the detector. The
multiplicity increases with HV for the whole surface area, but it is always less in the low-efficiency
zone in comparison with higher efficiency zones at each high voltage. The lowest resolutions for each
multiplicity, which is nearly diagonal points in these plots, are the regions of high purity signal. With high voltage there are relatively more events with a large avalanche pulse,
thus resolution with multiplicity four is reduced
with HV. Though the resolutions for multiplicity more than four are always larger than the events with multiplicity 1 to 3, except
for multiplicity one with very high voltage. The larger resolution is mainly due to the contamination of
streamer pulses. Similarly, for five-hit multiplicity, it is similar but happening at an even higher
voltage. At larger HV, e.g., 10.6\,V, events with multiplicity one is dominated by the region of
dead strips and consequently observed large resolution.
Overall, the observed position resolution for the events with four hits or more shows a substantial
deterioration in comparison with the events with multiplicities upto three.

\section{Efficiency using Selected Events}
\label{chap_efficiency}
\begin{figure}
  \center
  \includegraphics[width=0.66\linewidth]{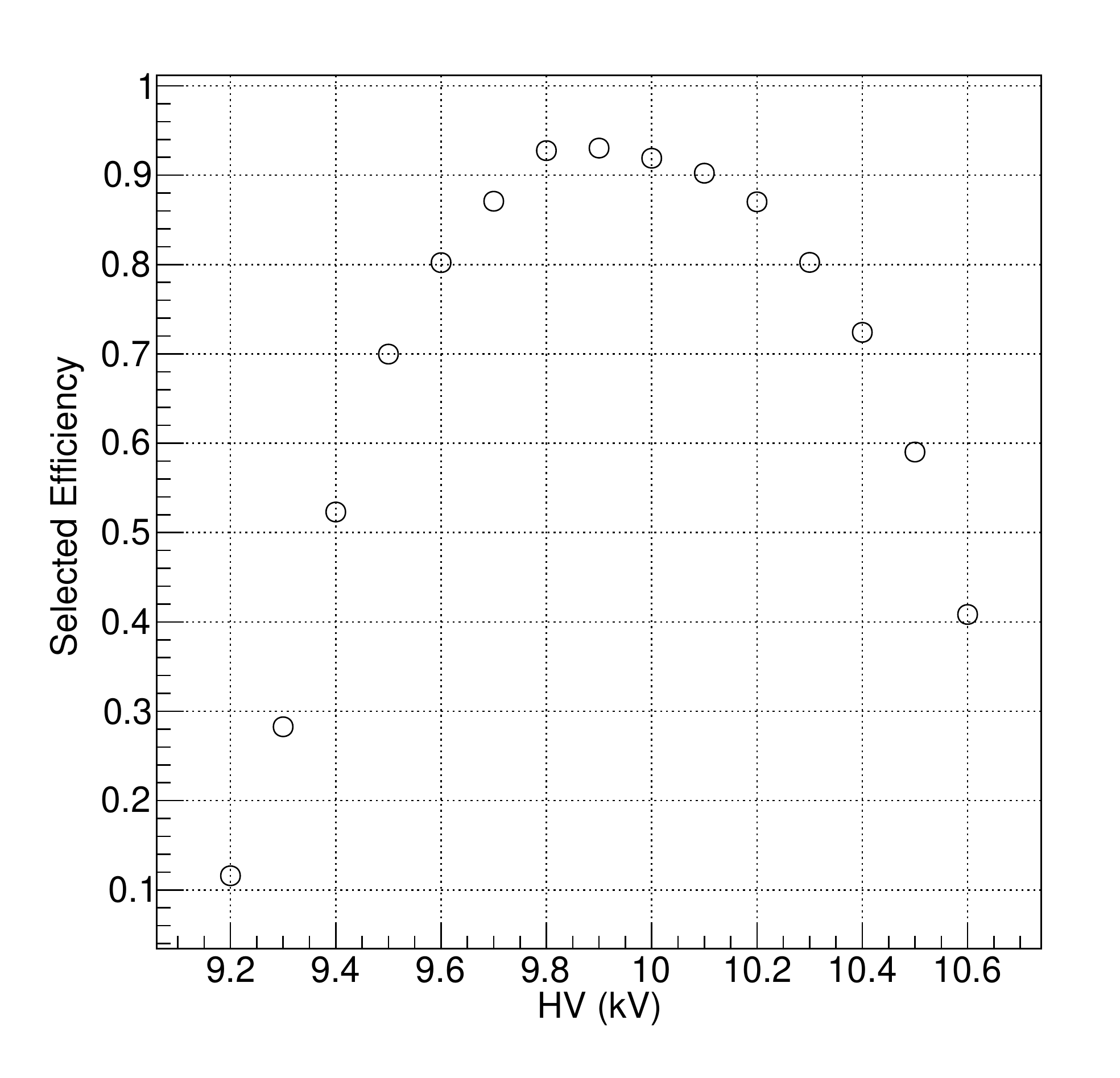}
  \caption{The measured efficiency from the selected events vs high voltage for the RPC under observation.}
  \label{fig:seleffhv}
\end{figure}

The efficiency of the RPC which was discussed in the section \ref{chap_conventional} is estimated without
any selection criteria on the multiplicities, the event can have any number of hits in the RPC
under the study. It was discussed in the previous section that the avalanche can induce
upto $\sim$ three strip hits and beyond that is not recommended to use for any physics
study due to worse localisation of signal. Thus, the definition of the
efficiency is modified with the criteria of accepting events with at most three consecutive hit
in this test RPC. The average efficiencies of the selected events defined as selected
efficiency for the whole detector as a function of applied voltage are shown in Fig.~\ref{fig:seleffhv}

The average efficiency has a peak at $\sim$9.8-9.9\,kV for the RPC under observation
suggests that the optimum operating voltage of the RPC under the study
should be 9.8-9.9\,kV.
Thus from the efficiency point of view, the optimum operating voltage of this RPC is less than
10.1\,kV which was obtained earlier with the simple efficiency measurement. Beyond 10.1\,kV, the
efficiency of the selected events are decreasing because of the rejection of events with large
multiplicities, mainly due to the formation of streamer pulses and
it is more for high gain regions.

\section{Position and Time resolution of the RPCs}
\label{chap_resolution}
\begin{figure}
  \center
  \includegraphics[width=1.1\linewidth]{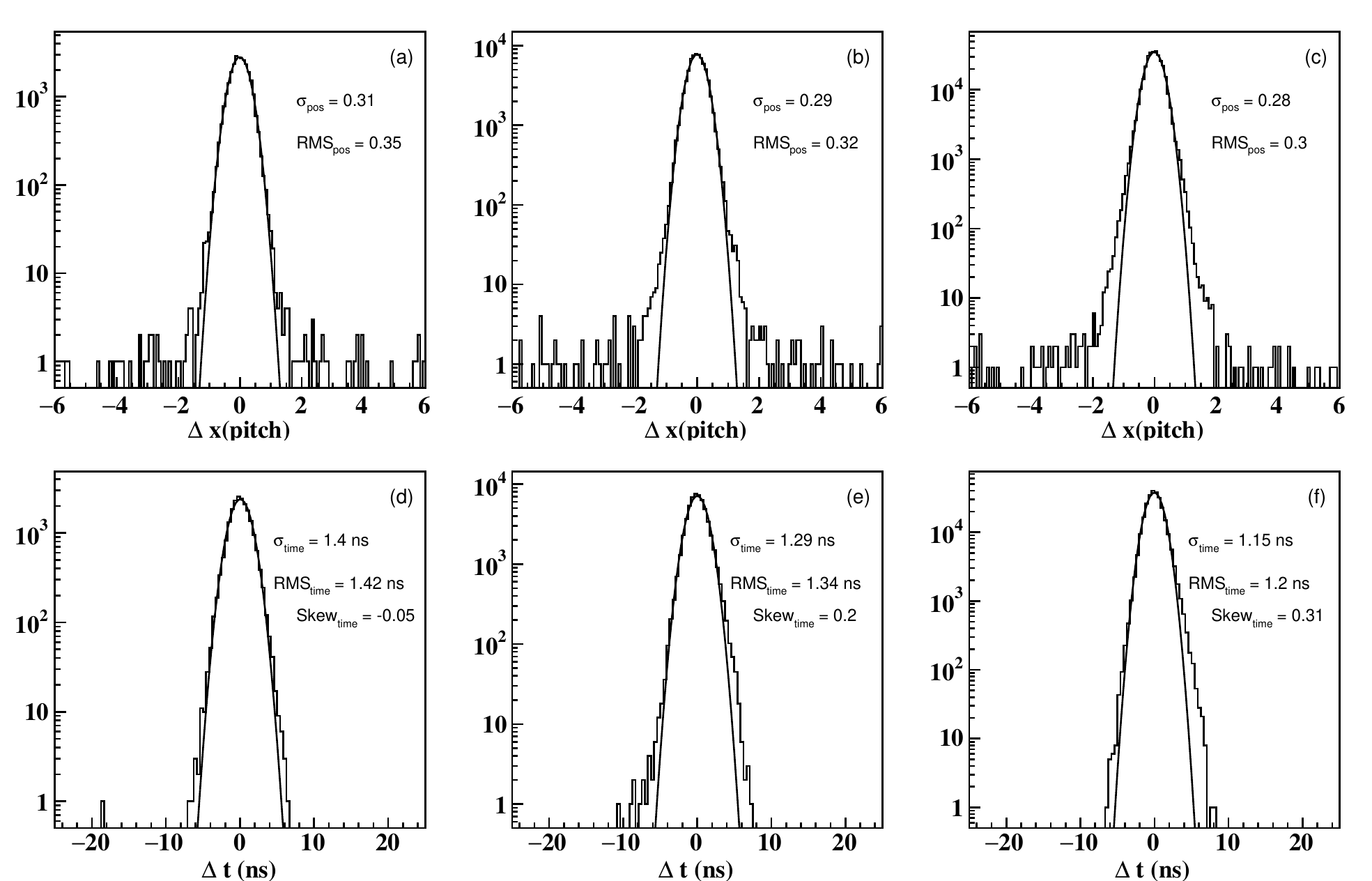}
  \caption{(a), (b) and (c) are the position residue distribution at
    9.8\,kV for efficiency zones of 20-30\,$\%$, 50-60\,$\%$ and
    80-90\,$\%$ respectively. (d), (e) and (f) are the time residue
    distribution at 9.8\,kV for efficiency zones of 20-30\,$\%$,
    50-60\,$\%$ and 80-90\,$\%$ respectively.}
  \label{fig:postimereso}
\end{figure}

The efficiency is not the only ultimate test for the optimisation of the applied HV. The position 
resolution drastically deteriorated for multiplicity four or more, but also these are worse
for multiplicity three in comparison with multiplicity one or two. Even after rejecting the events with hit multiplicities four or more, the physics parameters can be affected bit by the hit multiplicity three.
Until now the main parameter considered only the efficiency (with and without multiplicity
criteria) to choose the operating voltage of the RPC. But to look at more detailed performance
the physics parameters of the RPCs such as position and time resolution of the RPC
are very important. The position residues, the difference between the observed and extrapolated
positions are distributed and the resulting distribution follows the gaussian shape. The
observed distribution is fitted with the Gaussian function the fitted $\sigma_{pos}$ is considered
as the position resolution of the RPC. Similar to the position resolution, the time resolution is also
estimated by fitting the distribution of the difference between the observed and extrapolated
time of the RPC layer. The distribution of position and time residues at 9.8\,kV are shown in
Figs.~\ref{fig:postimereso} (a) to (f) for different efficiency
regions such as 20-30\,$\%$, 50-60\,$\%$ and 80-90\,$\%$.

\begin{figure}
  \center
  \includegraphics[width=0.99\linewidth]{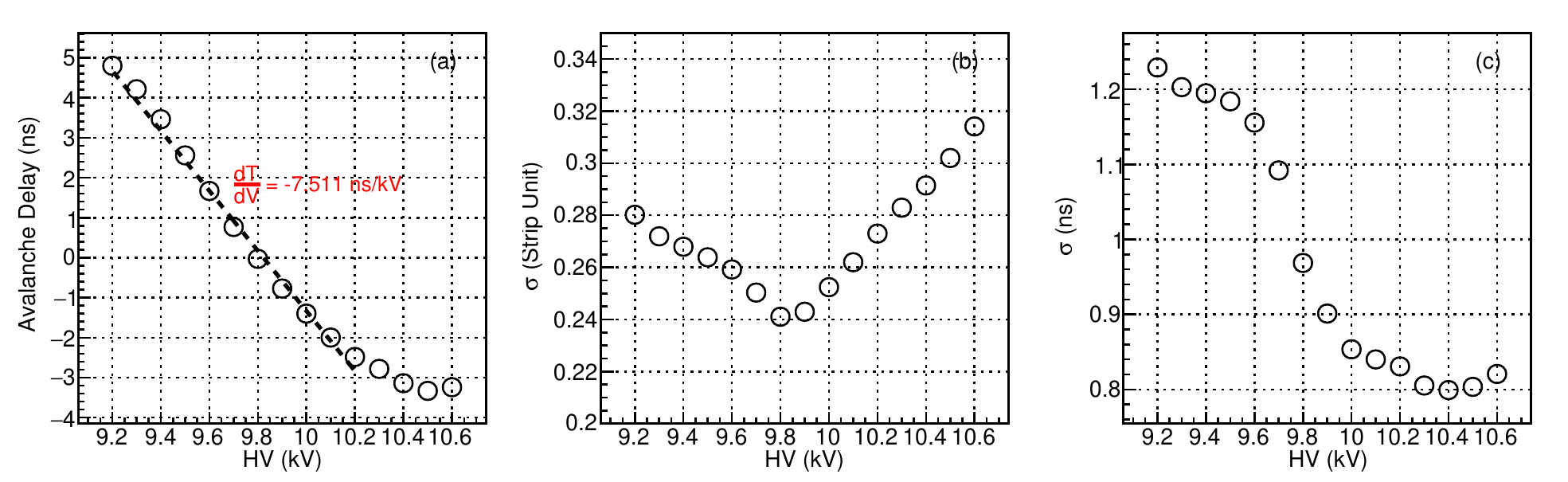}
  \caption{(a) is the mean of the time residues
    as a function of applied HV. (b) and (c) are the
    position and time resolution as a function of the applied voltage.}
  \label{fig:postimereshv}
\end{figure}

\begin{figure}
  \center
  \includegraphics[width=0.99\linewidth]{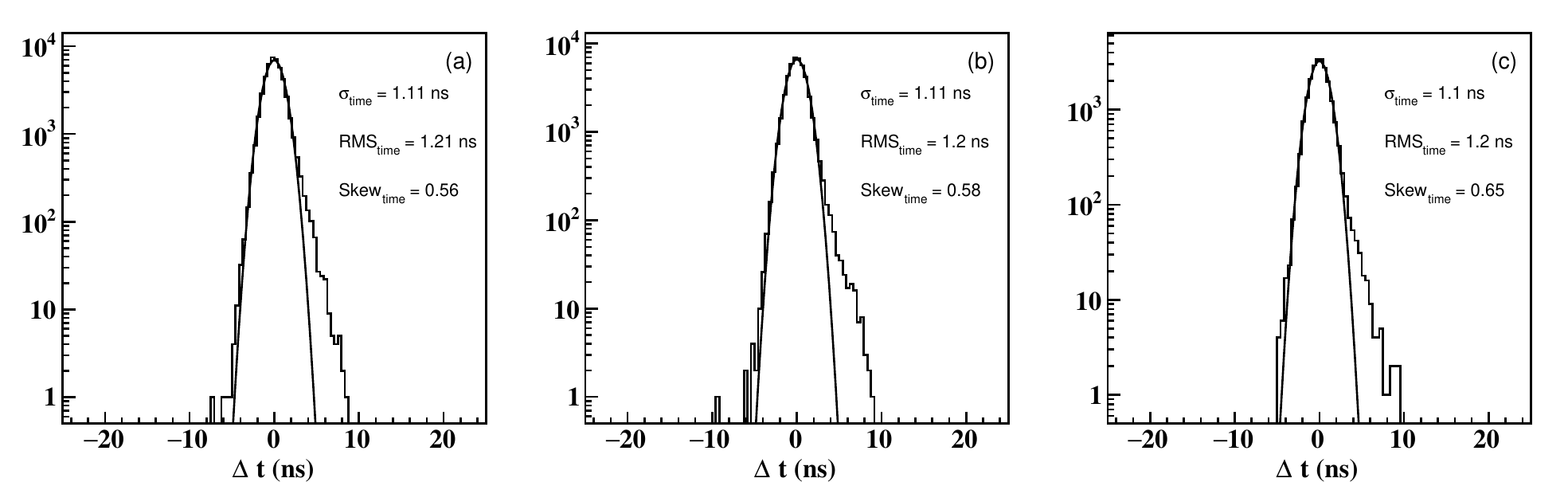}
  \caption{(a), (b) and (c) are the time residue distribution for the efficiency zone of 50-60\,$\%$ for 10.2, 10.4 and 10.6\,kV.}
  \label{fig:timeskew}
\end{figure}

The position and time offsets calculated from the muon data recorded
with different HV.
The mean of the time residues (can
be called avalanche delay with respect to the trigger signal) is expected to
show the correlation with HV (which is calculated with the time corrections calculated using the data from 9.8\,kV). As the HV increases the signal height
increases at a faster speed, consequently the threshold crossing time for the avalanche signal is earliar with respect to the trigger signal. From Fig.~\ref{fig:postimereshv}(a), the avalanche
delay vs HV is fitted using the straight-line upto 10.2\,kV until
there the linearity exists. The obtained slope for one of the planes is $-$7.51\,ns/kV.
The change in the slope beyond 10.2\,kV is mainly due to the effect of a large noise rate.
The fitted $\sigma_{pos}$ and $\sigma_{time}$ as a function of the
applied voltage for the RPC under study is shown in
Figs.~\ref{fig:postimereshv} (b) and (c).
The variation in the position resolution with different applied voltage shows improving behaviour
upto HV$_{pos}^{min}$=9.8-9.9\,kV. But beyond the HV$_{pos}^{min}$, the position resolution is
deteriorating due to the increased contribution of the multiplicities which
results in the poor localisation of the signal. The time resolution of the RPC under the study
improves as the applied voltage increases upto HV$_{time}^{min}$=10.4\,kV. The larger applied voltage
above HV$_{time}^{min}$ shows an increase in the time resolution.
During each avalanche, even for the noise pulse, the effective high voltage at the avalanche
position (few mm$^2$) is reduced below the threshold voltage to form an avalanche signal,
which causes the inefficiency as seen in the inset plot of Fig.~\ref{fig:effvshvrange}(a) and
gradually the high voltage is built up in that region. During the recovery time, the time taken
by the system to regain its stable voltage configuration, the effective HV is low and consequently
the gain is also low.
This explains the change in slope in Figs.~\ref{fig:postimereshv} (a) and increase of resolution in Figs.~\ref{fig:postimereshv} (c).
This argument is supported by the time residues distribution in the larger HV samples. Figs~\ref{fig:timeskew} (a), (b) and (c) are the time residue distribution for the efficiency zone of 50-60\,$\%$ for 10.2\,kV, 10.4\,kV and 10.6\,kV respectively. The time residue distribution for larger HV has an increasing number of events in the right side bump (which is small signals that cross the discriminator threshold later than the muon trigger time) and has increasing skewness at a larger voltage in comparison of skewness in Fig.~\ref{fig:postimereso}(d) to Fig.~\ref{fig:postimereso}(f).

The results of selected efficiency, as well as the position resolution, suggests that
the optimum operating point of this RPC is 9.8-9.9\,kV, not the 10.1-10.2\,kV, which was obtained using the
standard efficiency plots.

\begin{figure}
  \center
  \includegraphics[width=0.99\linewidth]{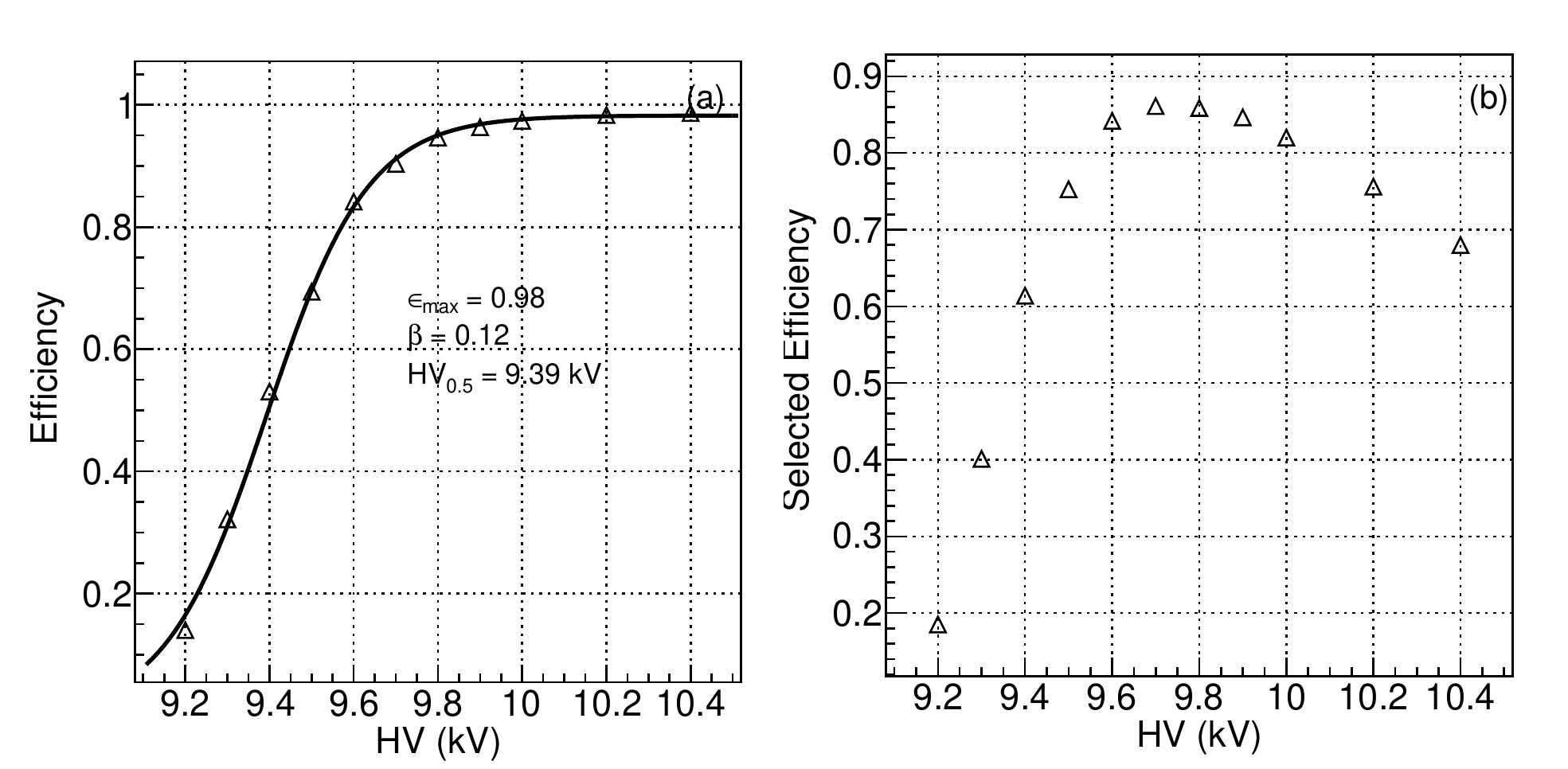}
  \caption{(a) and (b) are the default efficiency and selection
    efficiency as a function of applied HV for another RPC.}
  \label{fig:newrpcomb}
\end{figure}

\begin{figure}
  \center
  \includegraphics[width=0.99\linewidth]{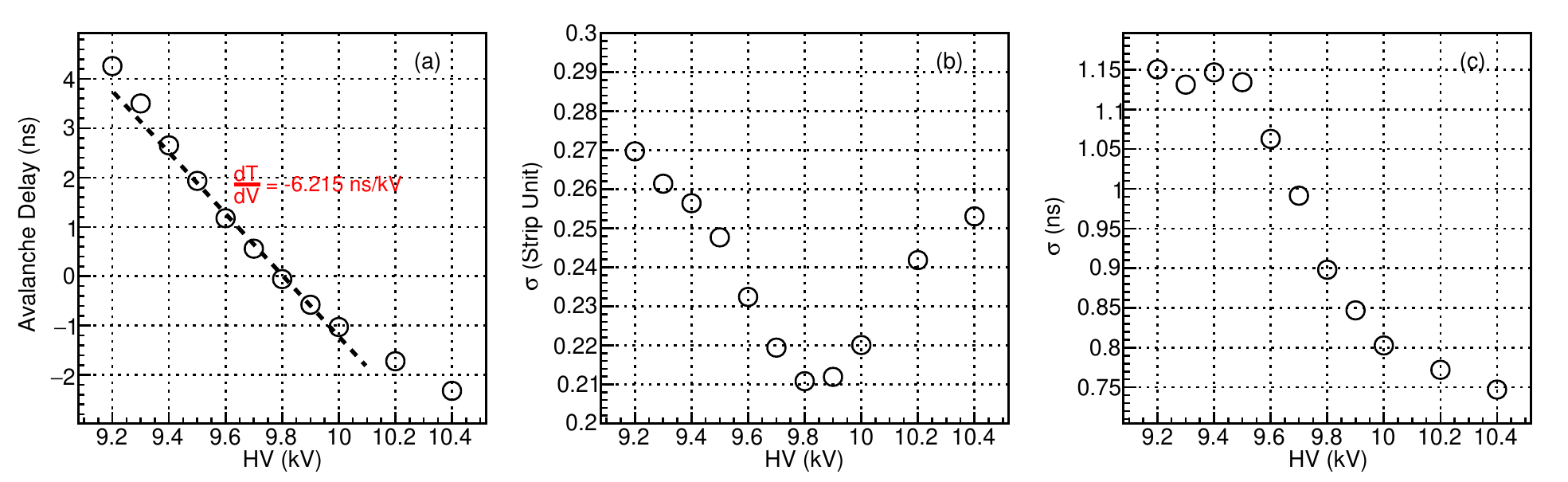}
  \caption{(a) is the mean of the time residues
    as a function of applied HV. (b) and (c) are the position and time resolution as a function of the applied voltage.}
  \label{fig:newpostimerms}
\end{figure}

Another RPC, of non-uniform gain, is also studied in the same
manner. The default efficiency and selected efficiency are shown in
Figs.~\ref{fig:newrpcomb}(a) and (b). Fig.~\ref{fig:newpostimerms}(a) is the mean of the time
residue as a function of HV is fitted using a straight line and the obtained slope for one of the planes is -6.21\,ns/kV. The
position resolution and time resolution at different operating voltages are shown in
Figs.~\ref{fig:newpostimerms}(b) and (c) respectively. That also shows that the optimum operating HV of the RPC for the best
physics performance is obtained at $\sim$9.8\,kV, not 10.1-10.2\,kV, which is obtained from the plateau of default efficiency.

\section{Conclusion}
\label{chap_conclusion}  
Two RPCs were tested at different operating HV to choose the optimum operating HV for the
best physics performance of the system. Due to the non-uniform response of these RPCs over
the whole surface area, different regions show the saturation of efficiencies at different
HV. Different zones can be considered as different individual RPCs, which should have
different optimum operating HV. Though the plateau of efficiency apparently shows
a larger operating voltage in terms of uniform gain, the optimum operating HV for the best
performance of the detector is lower than that. This paper shows examples of two RPCs with
large non-uniformities in gain. In general, these kinds of chambers is not installed in an
experiment. But, even one installs a good RPC in beginning, during the operational phase,
characteristics of the RPC might change, e.g. change in resistivity in the graphite coating,
bulging due to button popup etc., and then certainly one needs to check the optimum
operating HV for the best physics out of it, not simple efficiency plateau. Even with a small
variation of gain in an RPC, one should choose the operating HV by optimising the selected
efficiency, resolutions but not only overall efficiency.

\acknowledgments

We would also like to thank all members of the INO collaboration for their valuable inputs.



\end{document}